\documentclass[12pt]{article}
\newcommand{\nc}{\newcommand}
\nc{\rnc}{\renewcommand}
\rnc{\baselinestretch}{1.25}
\rnc{\arraystretch}{0.8}
\rnc{\thesection}{\arabic{section}.}
\rnc{\thesubsection}{\arabic{section}.\arabic{subsection}}
\rnc{\theequation}{\arabic{section}.\arabic{equation}}
\rnc{\thefootnote}{\dag}
\nc{\sect}[1]{\section{#1}\setcounter{equation}{0}}
\nc{\sub}[1]{\subsection{#1}}
\nc{\subsub}[1]{\subsubsection{#1}}
\nc{\be}{\begin{equation}}
\nc{\ee}{\end{equation}}
\nc{\bc}{\begin{center}}
\nc{\ec}{\end{center}}
\nc{\bpic}{\begin{picture}}
\nc{\epic}{\end{picture}}
\nc{\ba}[1]{\begin{array}{@{}#1@{}}}
\nc{\ea}{\end{array}}
\nc{\bea}{\begin{eqnarray}}
\nc{\eea}{\end{eqnarray}}
\nc{\el}[1]{\label{#1}}
\nc{\er}[1]{(\ref{#1})}

\nc{\itm}[1]{\\\noindent$\bullet$\ \ #1}
\nc{\ds}{\displaystyle}
\nc{\ts}{\textstyle}
\rnc{\ss}{\scriptstyle}
\nc{\sss}{\scriptscriptstyle}
\nc{\ru}[1]{\rule[-#1ex]{0ex}{#1ex}}
\rnc{\vec}[1]{\mbox{\boldmath$#1$}}
\font\tenmsb=msbm10 scaled \magstep1
\font\sevenmsb=msbm7 scaled \magstep1
\font\fivemsb=msbm5 scaled \magstep1
\newfam\msbfam
\textfont\msbfam=\tenmsb
\scriptfont\msbfam=\sevenmsb
\scriptscriptfont\msbfam=\fivemsb
\def\Bbb#1{{\fam\msbfam\relax#1}}
\nc{\p}[2]{\makebox(0,0)[#1]{$#2$}}
\nc{\pp}[2]{\makebox(0,0)[#1]{$\ss#2$}}
\nc{\ppp}[2]{\makebox(0,0)[#1]{$\sss#2$}}
\nc{\text}[6]{\begin{picture}(#1,#2)\put(#3,#4){\p{#5}{\ds#6}}\end{picture}}
\rnc{\a}{\bar{a}}
\nc{\A}{{\cal A}}
\nc{\B}{{\cal B}}
\nc{\C}{\Bbb C}
\rnc{\d}{\delta}
\nc{\D}{{\cal D}}
\nc{\e}{\varepsilon}
\nc{\eb}{\bar{\varepsilon}}
\nc{\G}{{\cal G}}
\nc{\hf}{{\sss1\!/\!2}}
\nc{\N}{{\cal N}}
\nc{\NN}{\Bbb N}
\rnc{\l}{\lambda}
\nc{\mi}{\!-\!}
\nc{\mipl}{\!\mp\!}
\nc{\pl}{\!+\!}
\nc{\plmi}{\!\pm\!}
\nc{\s}{\sigma}
\nc{\T}{{\cal T}}
\nc{\Z}{\Bbb Z}
\nc{\W}[5]{W\!\left(\,\begin{array}{@{}cc|@{\:}}#4&#3\\#1&#2\end{array}\;#5\right)}
\nc{\Wsm}[5]{W\!\Bigl(\:\ba{c@{}c}#4&#3\\#1&#2\ea\:\Big|\,#5\Bigr)}

\begin{document}
\bc
\vspace*{1mm}
\textbf{\Large Integrable Lattice Models for Conjugate $A^{(1)}_n$}\\
\bigskip\bigskip
{\large Roger E. Behrend and David E. Evans}\\
\bigskip
\textit{School of Mathematics,
Cardiff University,\\Cardiff, CF24 4YH, UK}\\
\smallskip
{\footnotesize\tt behrendr@cf.ac.uk, evansde@cf.ac.uk}\\
\begin{abstract}
\noindent A new class of $A^{(1)}_n$ integrable lattice models is presented.
These are interact-ion-round-a-face models based on fundamental nimrep graphs
associated with the $A^{(1)}_n$ conjugate modular invariants, there being
a model for each value of the rank and level.  The Boltzmann weights are
parameterized by elliptic theta functions and satisfy the Yang-Baxter
equation for any fixed value of the elliptic nome~$q$.
At $q=0$, the models provide representations of the Hecke algebra
and are expected to lead in the continuum limit to coset conformal field
theories related to
the $A^{(1)}_n$ conjugate modular invariants.
\end{abstract}
\ec
\bigskip
\sect{Introduction}
Modular invariants and nimreps
constitute important combinatorial objects in the classification
of rational conformal field theories.
(For an account of these objects, see for example \cite{Gan01,Gan02}).
However, a rational conformal field theory is not completely described by its
modular invariant and associated nimrep,
and in some cases much additional information can be provided by a corresponding
integrable lattice model.

This issue was considered for the case of
$A^{(1)}_n$ level-$k$ in~\cite{DifZub90a,DifZub90b,Dif92}.
As outlined there, for an $A^{(1)}_{n,k}$ modular invariant $M$ and
associated nimrep $\N$
there may exist a critical integrable interaction-round-a-face lattice model
whose continuum limit provides a realization of the
$A^{(1)}_{n,k\mi1}\times A^{(1)}_{n,1}/A^{(1)}_{n,k}$ coset conformal
field theory with modular invariant $I_{n,k\mi1}\otimes I_{n,1}\otimes M$,
where $I_{n,k\mi1}$ and $I_{n,1}$ are the $A^{(1)}_{n,k\mi1}$ and $A^{(1)}_{n,1}$
identity modular invariants.
The main features of such a model are that its spin states take values from the same set
as that which labels the rows and columns of the nimrep matrices $\N_\l$,
that the adjacency condition for spins on neighboring sites of the lattice
is given by the nimrep
graph for the fundamental vector representation of $A^{(1)}_{n,k}$, and that
the Boltzmann weights at some value of the spectral parameter
provide a representation of a certain quotient of the Hecke algebra.

For any modular data, two particularly simple and natural modular invariants are the identity
and the conjugate, with nimreps for the former being automatically given by
fusion matrices and nimreps for various cases of the
latter having recently been given explicitly
in~\cite{PetZub02,GabGan02}.
(Note, however, that if specialized characters are used, for which conjugate representations
have the same character, then the identity and conjugate modular invariants both give the same
modular invariant partition function.)
For the $A^{(1)}_{n,k}$ identity modular invariant, corresponding integrable
lattice models of the type outlined in
\cite{DifZub90a,DifZub90b,Dif92} are given in~\cite{JimMiwOka87a,JimMiwOka88a}.
The Boltzmann weights in these models
are parameterized by elliptic theta functions of
fixed nome~$q$, and it is then specifically the case $q=0$ which corresponds to criticality
and which gives models whose continuum limit leads to
coset conformal field theories.

In this paper we present general models for the case of the
$A^{(1)}_{n,k}$ conjugate modular invariant.  These models are
based, therefore, on the fundamental nimrep graphs of~\cite{PetZub02,GabGan02},
and as with the models of~\cite{JimMiwOka87a,JimMiwOka88a}
their Boltzmann weights are parameterized by
elliptic functions of nome $q$, with $q=0$ giving models
of the type outlined in~\cite{DifZub90a,DifZub90b,Dif92}.
We describe the nimrep graphs in Section~2, present and study the
Boltzmann weights for arbitrary $q$ in Section~3, and study the Boltzmann
weights for $q=0$ in Section~4.  In Section~5 we briefly
discuss some general matters arising
from the previous results, including the possibility of relationships
between the conjugate $A^{(1)}_{n,k}$ Boltzmann weights
and the $B^{(1)}_l$, $C^{(1)}_l$ and $A^{(2)}_l$
Boltzmann weights of~\cite{JimMiwOka88b,Kun91},
the existence of cells which intertwine identity and conjugate
$A^{(1)}_{n,k}$ Boltzmann weights
and the existence of fused conjugate $A^{(1)}_{n,k}$ models.

\sect{Graphs for Conjugate $A^{(1)}_n$}
In this section we explicitly describe the required nimrep matrices and graphs
associated with the $A^{(1)}_n$ conjugate modular invariants.  This material
is entirely based on the results of~\cite{PetZub02,GabGan02}, in which the nimrep matrices
were given in terms of their eigenvalues and eigenvectors, and also
in terms of certain $A^{(1)}_l$, $B^{(1)}_l$ and $C^{(1)}_l$ fusion matrices.

We are considering the affine Kac-Moody algebra $A^{(1)}_n=\widehat{sl}(n\pl1)$
of rank $n$ at fixed level $k$.
We shall from now on work with the shifted level, or altitude,
\be g\,=\,n\pl k\pl1\,,\ee
and denote $A^{(1)}_n$ at shifted level $g$ as $A_{n,g}$.
The integrable representations of $A_{n,g}$ are labeled by elements or weights
of the Weyl alcove
\be P_{n,g}\;=\;\{\l\equiv(\l_1,\ldots,\l_n)\in\NN^n\mid\l_1\pl\ldots\pl\l_n\!\le g\mi1\}\,.\ee
In this notation, the identity or vacuum representation is labelled by $(1,\ldots,1)$
and the $n$ fundamental representations by $(2,1,\ldots,1)$,
$(1,2,1,\ldots,1)$, \ldots, $(1,\ldots,1,2)$.  The fundamental representation
labeled by $(2,1,\ldots,1)$ is referred to as the vector representation and corresponds
to a Young diagram comprising a single square.
The cardinality of $P_{n,g}$ is $\Bigl(\ba{c}g\mi1\\n\ea\Bigr)$.
Conjugation of representations is given by
\be(\l_1,\dots,\l_n)^*=(\l_n,\dots,\l_1)\,.\ee
The fusion rule coefficients are denoted $N_{\l\mu}{}^\nu$ $(\l,\mu,\nu\in P_{n,g})$,
and lead to the fusion matrices $(N_\l)_{\mu\nu}=N_{\l\mu}{}^\nu$ which form a representation of
the fusion ring,
\be N_\l\;N_\mu\;=\;\sum_{\nu\in P_{n,g}}N_{\l\mu}{}^\nu\;N_\nu\,.\ee
A nimrep is any further collection of nonnegative-integer-entry matrices $\N_\l$
$(\l\in P_{n,g})$,
each with rows and columns labeled by the elements of some set $\B$, which
form a representation of the fusion ring,
\be\N_\l\;\N_\mu\;=\;\sum_{\nu\in P_{n,g}}N_{\l\mu}{}^\nu\;\N_\nu\,,\ee
and satisfy $\N_{(1,\ldots,1\!)}=I$ and $\N_{\l^*}=\N_\l{}^t$.
It is useful to regard nimrep matrices
as the adjacency matrices of graphs; that is, the vertices of such a graph
are the elements of $\B$ and there are $\N_{\l\,ab}$ edges connecting
vertex $a$ to vertex $b$.
In cases of interest in conformal field theory, it is expected
that each $A_{n,g}$ modular-invariant
matrix $M$ can be associated with at least one nimrep, with
the cardinalities of the corresponding $\B$ given by
$\mathrm{tr}M$.

Of sole interest here are the nimreps related to $A_{n,g}$ conjugation, as studied in detail
in \cite{PetZub02,GabGan02}.  These are associated with the modular-invariant matrix
$M_{\mu\nu}=\delta_{\mu,\nu*}$, and $|\B|$ is therefore
the number of self-conjugate representations in $P_{n,g}$.
Each such nimrep matrix is symmetric. (The eigenvalues of $\N_{\l}$ are
$S_{\l\mu}/S_{\l,(1\ldots1)}$ for all self-conjugate $\mu$, where $S$ is the
$A_{n,g}$ $S$-matrix, and from the conjugation properties of $S$ these are
seen to be real.)

The labeling sets for these $A_{n,g}$ conjugation nimreps are taken as
\be\el{B}\ba{l}\ru{2.7}\B\:=\\\;\;\left\{\ba{l}\ru{3}
\Bigl\{a\equiv(a_1,\ldots,a_{\frac{n\pl1}{2}})\in\NN^{\frac{n\pl1}{2}}\Bigm|
a_1\pl2(a_2\ldots\pl a_{\frac{n\pl1}{2}})\le g\mi1\Bigr\}\,,\;\:n\mbox{ odd}\\
\Bigl\{a\equiv(a_1,\ldots,a_{\frac{n}{2}})\in\NN^{\frac{n}{2}}\Bigm|
2(a_1\pl\ldots\pl a_{\frac{n}{2}\mi1})\pl a_{\frac{n}{2}}\le g\mi1,\;
(-1)^{g\,+\,a_{\!\frac{n}{2}}}=1\Bigr\}\,,\;n\mbox{ even.}\ea\right.\ea\ee
For $n=1$ and $2$, \er{B} should be interpreted as
\be\B\,=\,\left\{\ba{l}
\ru{2.4}\{1,2,\ldots,g\mi1\};\;\;n=1\\
\ru{2.4}\{1,3,\ldots,g\mi2\};\;\;n=2,\;g\mbox{ odd}\\
\{2,4,\ldots,g\mi2\};\;\;n=2,\;g\mbox{ even.}\ea\right.\ee
The labeling sets can also be generated using $P_{n,g}$ as
\be\B\,=\,\left\{\ba{l}
\ru{2.7}\Bigl\{(2a_1,a_2,\ldots,a_{\frac{n\pl1}{2}})\Bigm|
(a_1,\ldots,a_{\frac{n\pl1}{2}})\in P_{\frac{n\pl1}{2},\lfloor\frac{g+1}{2}\rfloor}\Bigr\}
\;\bigcup\\
\ru{3.8}\qquad\Bigl\{(2a_1\mi1,a_2,\ldots,a_{\frac{n\pl1}{2}})\Bigm|
(a_1,\ldots,a_{\frac{n\pl1}{2}})\in P_{\frac{n\pl1}{2},\lfloor\frac{g}{2}\rfloor\pl1}\Bigr\}
\,;\;\:n\mbox{ odd}\\
\ru{3.8}\Bigl\{(a_1,\ldots,a_{\frac{n}{2}\mi1},2a_{\frac{n}{2}}\mi1)\Bigm|
(a_1,\ldots,a_{\frac{n}{2}})\in P_{\frac{n}{2},\frac{g+1}{2}}\Bigr\}\,;\;\:
n\mbox{ even},\;g\mbox{ odd}\\
\Bigl\{(a_1,\ldots,a_{\frac{n}{2}\mi1},2a_{\frac{n}{2}})\Bigm|
(a_1,\ldots,a_{\frac{n}{2}})\in P_{\frac{n}{2},\frac{g}{2}}\Bigr\}\,;\;\:
n\mbox{ even},\;g\mbox{ even,}\ea\right.\ee
from which it follows that
$|\B|=\Biggl(\ba{c}\lfloor\frac{g-1}{2}\rfloor\ru{1.4}\\\frac{n+\!1}{2}\ea\Biggr)
+\Biggl(\ba{c}\lfloor\frac{g}{2}\rfloor\ru{1.4}\\\frac{n+\!1}{2}\ea\Biggr)$ for $n$ odd,
and $|\B|=\Biggl(\ba{c}\lfloor\frac{g-1}{2}\rfloor\ru{1.4}\\\frac{n}{2}\ea\Biggr)$
for $n$ even.\rule{0ex}{2.5ex}
We note that the labeling used here differs from that of \cite{GabGan02}
for the case $n$ even with $g$ odd, but is otherwise essentially equivalent to that
of \cite{PetZub02} and \cite{GabGan02}.

Of primary interest here are the nimrep matrices for the fundamental vector
representation $(2,1,\ldots,1)$, whose entries are given by
\be\el{N}\N_{(2,1,\ldots,1)\,ab}\;=\;\left\{\ba{l}\ru{4}\ds\sum_{i=1}^{\frac{n\pl1}{2}}
\Bigl(\d_{a-b,\,\e_i}+\d_{b-a,\,\e_{i}}\Bigr),\;\;n\mbox{ odd}\\
\ds\sum_{i=1}^{\frac{n}{2}}
\Bigl(\d_{a-b,\,\e_i}+\d_{b-a,\,\e_{i}}\Bigr)\,+\,(1\mi\d_{a_{\!\frac{n}{2}}\!,1})\:\d_{a,b}
\,,\;\;n\mbox{ even,}\ea\right.\ee
where $\e_i\in\Z^{\lfloor\!\frac{n\pl1}{2}\!\rfloor}$ are defined as:
\be\ba{|@{\quad}c@{\quad}|@{\quad}c@{\quad}|}
\hline
\rule[-1.6ex]{0ex}{4.8ex}n\mbox{ odd}&n\mbox{ even}\\
\hline
\rule[-1.8ex]{0ex}{5.1ex}\e_1=(1,0,\ldots,0)&\e_1=(1,0,\ldots,0)\\
\ru{1.8}\e_2=(-1,1,0,\ldots,0)&\e_2=(-1,1,0,\ldots,0)\\
\ru{1.8}\e_3=(0,-1,1,0,\ldots,0)&\e_3=(0,-1,1,0,\ldots,0)\\
\vdots&\vdots\\
\ru{1.8}\e_{\frac{n\mi1}{2}}=(0,\ldots,0,-1,1,0)&\e_{\frac{n}{2}-\!1}=(0,\ldots,0,-1,1,0)\\
\ru{2}\e_{\frac{n\pl1}{2}}=(0,\ldots,0,-1,1)&\e_{\frac{n}{2}}=(0,\ldots,0,-1,2)\\
\hline\ea\ee
For $n=1$ we take $\e_1=1$ and for $n=2$ we take $\e_1=2$.
We shall also use $\e_i\in\Z^{\lfloor\!\frac{n\pl1}{2}\!\rfloor}$ with $i\le0$, defined by
\be\e_0\,=\,(0,\ldots,0)\;\;(n\mbox{ even})\,;\qquad\e_i\,=\,-\e_{\!-i}\,,\;i<0\,.\ee
We shall denote the graph corresponding to $\N_{(2,1,\ldots,1)}$
as $\G_{n,g}$, or simply $\G$, and its adjacency matrix as $G$.
Thus the set of vertices of $\G$ is $\B$, $G=\N_{(2,1,\ldots,1)}$ and
any edge of $\G$ lies along $\e_i$, for some $i$.  We see from~\er{N} that
$G$ is symmetric with each entry either $0$ or $1$, so that $\G$ is an unoriented graph
with only single edges.
We shall denote the set of $r$-step paths on $\G$ by $\G^r$,
\be\G^r\,=\,\Bigl\{(a_0,\ldots,a_{r})\in\B^{r\pl1}\Bigm|
\prod_{j=0}^{r-1}G_{a_ja_{j+1}}=1\Bigr\}\,.\ee
Some examples of $\G$ for low rank ($n\le4$)
or low level ($k\le3$) are given in the following figures, in which
brackets and commas are omitted from the vertex labels for simplicity, and
in which $\A_l$ and $\D_l$ are the $A$ and $D$ Dynkin diagrams
and $\T_l$ is the `tadpole' graph (in which an end vertex
of $\A_l$ becomes self-adjacent):
\setlength{\unitlength}{7.5mm}
\[\text{2.1}{1}{0}{0.5}{l}{\G_{1,g}\;=}
\bpic(5,1)\multiput(0,0.5)(3,0){2}{\line(1,0){2}}
\multiput(2.2,0.5)(0.4,0){2}{\line(1,0){0.2}}
\multiput(0,0.5)(1,0){2}{\pp{}{\bullet}}
\multiput(4,0.5)(1,0){2}{\pp{}{\bullet}}
\put(0,0.35){\ppp{t}{1}}\put(1,0.35){\ppp{t}{2}}
\put(4,0.4){\ppp{t}{g-2}}\put(5,0.4){\ppp{t}{g-1}}\epic
\text{2.4}{1}{2.4}{0.5}{r}{=\;\A_{g\mi1}}\]

\[\text{2.6}{2.4}{0}{1.2}{l}{\G_{2,g}\;=\;\left\{\rule{0ex}{9mm}\right.}\bpic(8,2.4)
\put(0,1.4){\bpic(5,1)\multiput(0,0.5)(3,0){2}{\line(1,0){2}}
\multiput(2.2,0.5)(0.4,0){2}{\line(1,0){0.2}}
\multiput(0,0.5)(1,0){2}{\pp{}{\bullet}}
\multiput(4,0.5)(1,0){2}{\pp{}{\bullet}}
\put(0,0.35){\ppp{t}{1}}\put(1,0.35){\ppp{t}{3}}
\put(4,0.4){\ppp{t}{g-4}}\put(5,0.4){\ppp{t}{g-2}}
\put(1,0.65){\circle{0.3}}
\multiput(4,0.65)(1,0){2}{\circle{0.3}}\epic
\text{3}{1}{0.5}{0.5}{l}{,\quad g\mbox{ odd}}}
\put(0,0){\bpic(5,1)\multiput(0,0.5)(3,0){2}{\line(1,0){2}}
\multiput(2.2,0.5)(0.4,0){2}{\line(1,0){0.2}}
\multiput(0,0.5)(1,0){2}{\pp{}{\bullet}}
\multiput(4,0.5)(1,0){2}{\pp{}{\bullet}}
\put(0,0.35){\ppp{t}{2}}\put(1,0.35){\ppp{t}{4}}
\put(4,0.4){\ppp{t}{g-4}}\put(5,0.4){\ppp{t}{g-2}}
\multiput(0,0.65)(1,0){2}{\circle{0.3}}
\multiput(4,0.65)(1,0){2}{\circle{0.3}}\epic
\text{3}{1}{0.5}{0.5}{l}{,\quad g\mbox{ even}}}\epic\]

\[\text{1.65}{2}{0}{1}{l}{\G_{3,4}=}\bpic(0.4,2)
\put(0.2,1){\pp{}{\bullet}}
\put(0.2,0.85){\ppp{t}{11}}\epic\qquad\;
\text{1.85}{2}{0}{1}{l}{\G_{3,5}=}\bpic(1,2)\put(0,1){\line(1,0){1}}
\multiput(0,1)(1,0){2}{\pp{}{\bullet}}
\put(0,0.85){\ppp{t}{11}}\put(1,0.85){\ppp{t}{21}}\epic\qquad\quad
\text{1.75}{2}{0}{1}{l}{\G_{3,6}=}\bpic(2,2)\put(0,0.5){\line(1,0){2}}
\put(1,0.5){\line(0,1){1}}
\multiput(0,0.5)(1,0){3}{\pp{}{\bullet}}\put(1,1.5){\pp{}{\bullet}}
\put(0,0.35){\ppp{t}{11}}\put(1,0.35){\ppp{t}{21}}\put(2,0.35){\ppp{t}{31}}
\put(1,1.65){\ppp{b}{12}}\epic\qquad\;
\text{1.75}{2}{0}{1}{l}{\G_{3,7}=}\bpic(3,2)\put(0,0.5){\line(1,0){3}}
\put(1,1.5){\line(1,0){1}}\multiput(1,0.5)(1,0){2}{\line(0,1){1}}
\multiput(0,0.5)(1,0){4}{\pp{}{\bullet}}
\multiput(1,1.5)(1,0){2}{\pp{}{\bullet}}
\put(0,0.35){\ppp{t}{11}}\put(1,0.35){\ppp{t}{21}}
\put(2,0.35){\ppp{t}{31}}\put(3,0.35){\ppp{t}{41}}
\put(0.95,1.58){\ppp{br}{12}}\put(2.05,1.58){\ppp{bl}{22}}\epic\]
\[\text{2}{3}{0}{1.5}{tl}{\G_{3,8}\,=}\bpic(4,3)\put(0,0.5){\line(1,0){4}}
\put(1,1.5){\line(1,0){2}}\put(2,0.5){\line(0,1){2}}
\multiput(1,0.5)(2,0){2}{\line(0,1){1}}\put(2,2.5){\pp{}{\bullet}}
\multiput(0,0.5)(1,0){5}{\pp{}{\bullet}}
\multiput(1,1.5)(1,0){3}{\pp{}{\bullet}}
\put(0,0.35){\ppp{t}{11}}\put(1,0.35){\ppp{t}{21}}
\put(2,0.35){\ppp{t}{31}}\put(3,0.35){\ppp{t}{41}}\put(4,0.35){\ppp{t}{51}}
\put(2,2.67){\ppp{b}{13}}
\put(0.95,1.58){\ppp{br}{12}}\put(3.05,1.58){\ppp{bl}{32}}
\put(2.08,1.58){\ppp{bl}{22}}\epic\qquad\qquad
\text{2}{3}{0}{1.5}{tl}{\G_{3,9}\,=}\bpic(5,3)\put(0,0.5){\line(1,0){5}}
\put(1,1.5){\line(1,0){3}}
\put(2,2.5){\line(1,0){1}}
\multiput(1,0.5)(3,0){2}{\line(0,1){1}}
\multiput(2,0.5)(1,0){2}{\line(0,1){2}}
\multiput(0,0.5)(1,0){6}{\pp{}{\bullet}}
\multiput(1,1.5)(1,0){4}{\pp{}{\bullet}}
\multiput(2,2.5)(1,0){2}{\pp{}{\bullet}}
\put(0,0.35){\ppp{t}{11}}\put(1,0.35){\ppp{t}{21}}
\put(2,0.35){\ppp{t}{31}}\put(3,0.35){\ppp{t}{41}}
\put(4,0.35){\ppp{t}{51}}\put(5,0.35){\ppp{t}{61}}
\put(1.9,2.65){\ppp{b}{13}}\put(3.1,2.65){\ppp{b}{23}}
\put(0.95,1.58){\ppp{br}{12}}\put(4.05,1.58){\ppp{bl}{42}}
\put(1.95,1.58){\ppp{br}{22}}\put(3.08,1.58){\ppp{bl}{32}}\epic\]

\[\text{0}{3}{0}{1.9}{l}{\G_{4,5}=}
\text{1.65}{3}{0.4}{0.8}{l}{\G_{4,6}=}\bpic(0.4,2)
\put(0.2,1.9){\pp{}{\bullet}}
\put(0.2,1.8){\ppp{t}{11}}
\put(0.6,0.8){\pp{}{\bullet}}\put(0.6,0.95){\circle{0.3}}
\put(0.6,0.7){\ppp{t}{12}}\epic\qquad
\text{1.7}{3}{0}{1.5}{tl}{\G_{4,7}=}\bpic(1,3)\put(0,1){\line(1,0){1}}
\put(1,1){\line(0,1){1}}
\multiput(0,1)(1,0){2}{\pp{}{\bullet}}
\put(1,2){\pp{}{\bullet}}
\put(1,2.15){\circle{0.3}}
\put(0,0.85){\ppp{t}{11}}\put(1,0.85){\ppp{t}{21}}
\put(1.12,2){\ppp{tl}{13}}
\epic\qquad
\text{1.8}{3}{0}{1.5}{tl}{\G_{4,8}=}\bpic(1,3)\put(0,1){\line(1,0){1}}
\put(1,1){\line(0,1){1}}
\multiput(0,1)(1,0){2}{\pp{}{\bullet}}
\put(1,2){\pp{}{\bullet}}
\multiput(0,1.15)(1,0){2}{\circle{0.3}}\put(1,2.15){\circle{0.3}}
\put(0,0.85){\ppp{t}{12}}\put(1,0.85){\ppp{t}{22}}
\put(1.12,2){\ppp{tl}{14}}
\epic\qquad
\text{1.6}{3}{0}{1.5}{tl}{\G_{4,9}=}\bpic(2,3)\put(0,0.5){\line(1,0){2}}
\put(1,1.5){\line(1,0){1}}\put(1,0.5){\line(0,1){1}}
\put(2,0.5){\line(0,1){2}}
\put(2,2.5){\pp{}{\bullet}}
\multiput(0,0.5)(1,0){3}{\pp{}{\bullet}}
\multiput(1,1.5)(1,0){2}{\pp{}{\bullet}}
\multiput(1,1.65)(1,0){2}{\circle{0.3}}\put(2,2.65){\circle{0.3}}
\put(0,0.35){\ppp{t}{11}}\put(1,0.35){\ppp{t}{21}}\put(2,0.35){\ppp{t}{31}}
\put(2.12,1.5){\ppp{tl}{23}}\put(0.92,1.5){\ppp{tr}{13}}
\put(2.1,2.5){\ppp{tl}{15}}\epic\qquad
\text{1.8}{3}{0}{1.5}{tl}{\G_{4,10}=}\bpic(2,3)\put(0,0.5){\line(1,0){2}}
\put(1,1.5){\line(1,0){1}}\put(1,0.5){\line(0,1){1}}
\put(2,0.5){\line(0,1){2}}
\put(2,2.5){\pp{}{\bullet}}
\multiput(0,0.5)(1,0){3}{\pp{}{\bullet}}
\multiput(1,1.5)(1,0){2}{\pp{}{\bullet}}
\multiput(0,0.65)(1,0){3}{\circle{0.3}}
\multiput(1,1.65)(1,0){2}{\circle{0.3}}\put(2,2.65){\circle{0.3}}
\put(0,0.35){\ppp{t}{12}}\put(1,0.35){\ppp{t}{22}}\put(2,0.35){\ppp{t}{32}}
\put(2.12,1.5){\ppp{tl}{24}}\put(0.92,1.5){\ppp{tr}{14}}
\put(2.1,2.5){\ppp{tl}{16}}\epic\]
\setlength{\unitlength}{10mm}
\[\text{1.75}{2}{0}{1}{l}{\G_{n,n\pl1}\,=}\bpic(0.4,2)
\put(0.2,1){\pp{}{\bullet}}
\put(0.2,0.9){\ppp{t}{1\ldots1}}\epic\qquad\qquad
\text{2.4}{2}{0}{1}{l}{\G_{n,n\pl2}\,=\,\left\{\rule{0ex}{8mm}\right.}
\bpic(1,2)\put(0,1.5){\line(1,0){1}}
\multiput(0,1.5)(1,0){2}{\pp{}{\bullet}}
\put(0,1.35){\ppp{t}{1\ldots1}}\put(1,1.35){\ppp{t}{21\ldots1}}
\put(0.5,0.5){\pp{}{\bullet}}\put(0.5,0.6){\circle{0.2}}
\put(0.5,0.4){\ppp{t}{1\ldots12}}
\put(1.1,0.5){\p{l}{,\quad n\mbox{ even}}}\epic
\text{3.3}{2}{3.3}{1.5}{r}{\approx\A_2\,,\quad n\mbox{ odd}}\]
\[\text{2.7}{2.5}{0}{1.4}{l}{\G_{n,n\pl3}\,=\,\left\{\rule{0ex}{13mm}\right.}\bpic(9.7,2.5)
\put(0,1.4){\bpic(6,1)
\put(1,0.5){\line(-2,1){1}}
\put(1,0.5){\line(-2,-1){1}}
\put(1,0.5){\line(1,0){2}}\put(4,0.5){\line(1,0){2}}
\multiput(3.2,0.5)(0.4,0){2}{\line(1,0){0.2}}
\multiput(0,0)(0,1){2}{\pp{}{\bullet}}
\multiput(1,0.5)(1,0){2}{\pp{}{\bullet}}
\multiput(5,0.5)(1,0){2}{\pp{}{\bullet}}
\put(-0.05,-0.02){\ppp{t}{31\ldots1}}\put(-0.1,1.1){\ppp{b}{1\ldots1}}
\put(1.25,0.35){\ppp{t}{21\ldots1}}
\put(2.28,0.35){\ppp{t}{121\ldots1}}
\put(4.9,0.35){\ppp{t}{1\ldots121}}\put(6.1,0.35){\ppp{t}{1\ldots12}}\epic
\text{3.7}{1}{0.7}{0.5}{l}{\approx\;\D_{\!\frac{n\pl5}{2}}\,,\quad n\mbox{ odd}}}
\put(0,0){\bpic(6,1)
\put(0,0.5){\line(1,0){3}}
\put(4,0.5){\line(1,0){2}}
\multiput(3.2,0.5)(0.4,0){2}{\line(1,0){0.2}}
\multiput(0,0.5)(1,0){3}{\pp{}{\bullet}}
\multiput(5,0.5)(1,0){2}{\pp{}{\bullet}}
\put(-0.1,0.35){\ppp{t}{1\ldots1}}\put(1,0.35){\ppp{t}{21\ldots1}}
\put(2.2,0.35){\ppp{t}{121\ldots1}}
\put(4.9,0.35){\ppp{t}{1\ldots121}}\put(6.1,0.35){\ppp{t}{1\ldots13}}
\put(6,0.625){\circle{0.25}}\epic
\text{3.7}{1}{0.7}{0.5}{l}{\approx\;\T_{\frac{n}{2}+\!1}\,,\quad n\mbox{ even}}}\epic\]

\[\text{2.6}{2.9}{0}{1.4}{l}{\G_{n,n\pl4}\,=\,\left\{\rule{0ex}{14mm}\right.}\bpic(9.7,2.9)
\put(0,1.4){\bpic(6,1)
\multiput(1,0.3)(1,0){2}{\line(0,1){0.7}}
\multiput(5,0.3)(1,0){2}{\line(0,1){0.7}}
\put(0,0.3){\line(1,0){3}}\put(4,0.3){\line(1,0){2}}
\put(0,1){\line(1,0){3}}\put(4,1){\line(1,0){2}}
\multiput(3.2,0.3)(0.4,0){2}{\line(1,0){0.2}}
\multiput(0,0.3)(1,0){3}{\pp{}{\bullet}}
\multiput(5,0.3)(1,0){2}{\pp{}{\bullet}}
\multiput(3.2,1)(0.4,0){2}{\line(1,0){0.2}}
\multiput(0,1)(1,0){3}{\pp{}{\bullet}}
\multiput(5,1)(1,0){2}{\pp{}{\bullet}}
\put(-0.1,0.15){\ppp{t}{41\ldots1}}\put(1,0.15){\ppp{t}{31\ldots1}}
\put(2.2,0.15){\ppp{t}{221\ldots1}}
\put(4.9,0.15){\ppp{t}{21\ldots121}}\put(6.15,0.15){\ppp{t}{21\ldots12}}
\put(-0.1,1.15){\ppp{b}{1\ldots1}}\put(1,1.15){\ppp{b}{21\ldots1}}
\put(2.2,1.15){\ppp{b}{121\ldots1}}
\put(4.9,1.15){\ppp{b}{1\ldots121}}\put(6.1,1.15){\ppp{b}{1\ldots12}}\epic
\text{3.7}{1}{0.7}{0.5}{l}{,\quad n\mbox{ odd}}}
\put(0,0){\bpic(6,1)
\put(0,0.5){\line(1,0){3}}\put(4,0.5){\line(1,0){2}}
\multiput(3.2,0.5)(0.4,0){2}{\line(1,0){0.2}}
\multiput(0,0.5)(1,0){3}{\pp{}{\bullet}}
\multiput(5,0.5)(1,0){2}{\pp{}{\bullet}}
\put(-0.1,0.35){\ppp{t}{1\ldots12}}\put(1,0.35){\ppp{t}{21\ldots12}}
\put(2.2,0.35){\ppp{t}{121\ldots12}}
\put(4.9,0.35){\ppp{t}{1\ldots122}}\put(6.1,0.35){\ppp{t}{1\ldots14}}
\multiput(0,0.625)(1,0){3}{\circle{0.25}}
\multiput(5,0.625)(1,0){2}{\circle{0.25}}\epic
\text{3.7}{1}{0.7}{0.5}{l}{\approx\;\G_{2,n\pl4}\,,\quad n\mbox{ even}}}\epic\]
We note that $\B$ can be regarded as being a subset of the Weyl alcove
$P^C_{\frac{n\pl1}{2},g}$ for $n$ odd or $P^B_{\frac{n}{2},g}$ for $n$ even,
and that with this identification we have
\be\el{BC}\N_{(2,1,\ldots,1)\,ab}\:=\:\left\{\ba{l}
\ru{2.8}N^C_{\;(2,1,\ldots,1)\;a}{}^b\,,\;\;n\mbox{ odd}\\
N^B_{\;(2,1,\ldots,1)\;a}{}^b\,,\;\;n\mbox{ even},\ea\right.\ee
where $B$ and $C$ refer to the algebras $C^{(1)}_{\frac{n\pl1}{2}}$ and
$B^{(1)}_{\frac{n}{2}}$, each at shifted level $g$
(or levels $g\mi\frac{n\pl1}{2}\mi1$ and $g\mi n\pl1$ respectively),
and $N^C_{\,\;ab}{}^c$ and $N^B_{\,\;ab}{}^c$
are their fusion coefficients (with $(2,1,\ldots,1)$ on the LHS of \er{BC}
a weight of the $A_{n,g}$ Weyl alcove $P_{n,g}$,
and $(2,1,\ldots,1)$ on the RHS a
weight of the Weyl alcoves $P^C_{\frac{n\pl1}{2},g}$ or $P^B_{\frac{n}{2},g}$).

For the case $n$ odd and the case $n$ even with $g$ even, there exist involutions $J$ of $\B$,
\be\el{J}J(a_1,\ldots,a_{\lfloor\!\frac{n\pl1}{2}\!\rfloor})\,=\,\left\{\ba{l}
\ru{3}(g\mi a_1\mi2a_2\mi\ldots\mi2a_{\frac{n\pl1}{2}},a_2,\ldots,a_{\frac{n\pl1}{2}})\,;\;\;
n\mbox{ odd}\\
(a_{\frac{n}{2}\mi1},\ldots,a_1,g\mi2a_1\mi\ldots\mi 2a_{\frac{n}{2}\mi1}\mi a_{\frac{n}{2}})
\,;\;\;n\mbox{ even},\;g\mbox{ even},\ea\right.\ee
for which \be\N_{\l,\;Ja,\,Jb}\,=\,\N_{\l\;ab}\ee
for any $\l\in P_{n,g}$ and $a,b\in\B$.

In Section 3 it will be convenient to use orthogonal coordinates for each $a\in\B$,
defined as
\be\a_i=\left\{\ba{l}
\ru{4}\ds\sum_{j=i}^{\frac{n\pl1}{2}}\!a_j\,,\;n\mbox{ odd}\\
\ds\sum_{j=i}^{\frac{n}{2}-1}\!a_j\,+\,\frac{a_{\frac{n}{2}}}{2}
\,,\;n\mbox{ even}\ea\right.\!,\;i>0\,;\qquad\a_i=-\a_{-i}\,,\;i<0\,;
\qquad\a_0=\ts-\frac{1}{2}\,.\ee
We see from \er{N} that if $(a_1,\ldots,a_{\lfloor\!\frac{n\pl1}{2}\!\rfloor})$
is relabeled as $(\a_1,\ldots,\a_{\lfloor\!\frac{n\pl1}{2}\!\rfloor})$,
then the nonzero edges of $\G$ lie along
$\pm\eb_1,\ldots,\pm\eb_{\lfloor\!\frac{n\pl1}{2}\!\rfloor}$, where $(\eb_i)_j=\d_{ij}$.

\sect{Boltzmann Weights for Conjugate $A^{(1)}_n$}
In this section we present Boltzmann weights for integrable lattice models based
on the conjugate $A_{n,g}$ graphs $\G$.  In these models,
a spin is attached to each site of a
two-dimensional square lattice, with the possible states of each spin being
the vertices of $\G$ and there being a lattice adjacency condition stipulating
that, in any assignment of spin states to the lattice, the states on each pair
of nearest-neighbor sites must correspond to an edge of $\G$. These
models are interaction-round-a-face models, so that Boltzmann weights
$\Wsm{b\:\;}{c}{d}{a\:\;}{u}$ are associated with sets of four spin states,
$a$, $b$, $c$, $d$, adjacent around a square face, where $u\in\C$ is
the spectral parameter.
The partition function of the model is then the sum,
over all possible spin assignments, of
products of Boltzmann weights over all square faces of the lattice.
The Boltzmann weights introduced here satisfy various properties,
and in particular the Yang-Baxter equation which ensures the integrability of the
model.  The notation we use is similar to that
of~\cite{JimMiwOka87a,JimMiwOka88a,JimMiwOka88b,Kun91}.

We define
\be\el{ef}[u]_1=\vartheta_1\Bigl(\frac{u\pi}{g},\,q\Bigr)\,,\qquad
[u]_4=\vartheta_4\Bigl(\frac{u\pi}{g},\,q\Bigr),\ee
where $\vartheta_1$ and $\vartheta_4$ are, up to a factor of
$2q^{\sss1\!/\!4}$ in $\vartheta_1$, elliptic theta functions of fixed nome $q$,
with $|q|<1$,
\be\ba{l}\ds\ru{3}
\vartheta_1(u,q)\,=\,\sin\!u\,
\prod_{n=1}^{\infty}\bigl(1\mi2q^{2n}\cos\!2u\pl q^{4n}\bigr)
\bigl(1\mi q^{2n}\bigr)\\
\ds\vartheta_4(u,q)\,=\,
\prod_{n=1}^{\infty}\bigl(1\mi2q^{2n\mi1}\cos\!2u\pl q^{4n\mi2}\bigr)
\bigl(1\mi q^{2n}\bigr).\ea\ee
The Boltzmann weights are then given as
\be\el{W}\ba{l}
\ds\ru{5}\W{a\pl\e_i}{a\pl2\e_i}{a\pl\e_i}{a}{u}\:=\:\frac{[\l\mi u]_4\,[1\mi u]_1}
{[\l]_4\,[1]_1}\,,\;\;i\ne0\\
\ds\ru{5}\W{a\pl\e_i}{a\pl\e_i\pl\e_j}{a\pl\e_i}{a}{u}\:=\:
\frac{[\l\mi u]_4\,[\a_i\mi\a_j\pl u]_1}
{[\l]_4\,[\a_i\mi\a_j]_1}\,,\;\;i\ne\pm j\\
\ds\ru{5}\W{a\pl\e_i}{a\pl\e_i\pl\e_j}{a\pl\e_j}{a}{u}\:=\:
\left(\frac{[\a_i\mi\a_j\mi1]_1\,[\a_i\mi\a_j\pl1]_1}{[\a_i\mi\a_j]_1^{\:\;2}}\right)^{\!\!\!\hf}
\frac{[\l\mi u]_4\,[u]_1}{[\l]_4\,[1]_1}\,,\;\;i\ne\pm j\\
\ds\ru{5}\W{a\pl\e_i}{a}{a\pl\e_j}{a}{u}\:=\:
\s_{\!ij}\,(\mi1)^{n\pl1}(\psi_{a,i}\,\psi_{a,j})^{\!\hf}\,
\frac{[\a_i\pl\a_j\pl1\mi\l\pl u]_4\,[u]_1}{[\l]_4\,[\a_i\pl\a_j\pl1]_1}\,,\;\;i\ne j\\
\ds\ru{5}\W{a\pl\e_i}{a}{a\pl\e_i}{a}{u}\:=\:
\frac{[\l\mi u]_4\,[2\a_i\pl1\pl u]_1+
(\mi1)^{n\pl1}\,\psi_{\!a,i}\,[2\a_i\pl1\mi\l\pl u]_4\,[u]_1}{[\l]_4\,[2\a_i\pl1]_1}\\
\ds\qquad\qquad\qquad\qquad=\:\frac{[\l\pl u]_4\,[2\a_i\pl1\mi2\l\pl u]_1+
(\mi1)^n\,\phi_{a,i}\,[2\a_i\pl1\mi\l\pl u]_4\,[u]_1}{[\l]_4\,[2\a_i\pl1\mi2\l]_1}
\ea\ee
where
\be\el{l}\l\,=\,\frac{n\pl1}{2}\,,\qquad\qquad\s_{\!ij}\,=\,
\left\{\ba{l}\ru{1.7}\mathrm{sign}(i)\:\mathrm{sign}(j)\,,\;n\mbox{ odd}\\
1\,,\;n\mbox{ even,}\ea\right.\ee

\be\el{psiai1}\psi_{a,i}\,=\,\frac{\psi_{a+\e_i}}{\psi_a}\,,\qquad\qquad\ee
\be\el{psia}\psi_a\,=\,\left\{\ba{l}\ds\ru{4}\prod_{i=1}^{\frac{n\pl1}{2}}[\a_i]_4\,[\a_i]_1\;
\prod_{1\le i<j\le\frac{n\pl1}{2}}\![\a_i\mi\a_j]_1\,[\a_i\pl\a_j]_1\,,\;n\mbox{ odd}\\
\ds\prod_{i=1}^{\frac{n}{2}}\!\frac{[2\a_i]_1}{[\a_i]_4}\;
\prod_{1\le i<j\le\frac{n}{2}}\![\a_i\mi\a_j]_1\,[\a_i\pl\a_j]_1\,,\;\,n\mbox{ even,}
\ea\right.\ee
and
\be\phi_{a,i}\:=\!\sum_{\stackrel{\ru{0.7}\ss b\,=\,a+\e_j\in\,\B}
{\sss(G_{ab}=1,\;j\ne i)}}\!\!
\frac{[\a_i\pl\a_j\pl1\mi2\l]_1}{[\a_i\pl\a_j\pl1]_1}\:\psi_{a,j}\,.\ee
Substituting \er{psia} into \er{psiai1}, $\psi_{a,i}$ for $i\ne0$ can be written explicitly as
\be\el{psiai2}\psi_{a,i}\:=\:\left\{\ba{l}\ds\ru{4.6}
\frac{[\a_i\pl1]_4\,[\a_i\pl1]_1}{[\a_i]_4\,[\a_i]_1}\,
\prod_{j\in\{\pm\!1,\ldots,\pm\!\frac{n\pl1}{2}\}/\!\{\pm i\}}\!\!
\frac{[\a_i\pl\a_j\pl1]_1}{[\a_i\pl\a_j]_1}\,,\;\;n\mbox{ odd}\\
\ds\frac{[\a_i]_4\,[2\a_i\pl2]_1}{[\a_i\pl1]_4\,[2\a_i]_1}\,
\prod_{j\in\{\pm\!1,\ldots,\pm\!\frac{n}{2}\}/\!\{\pm i\}}\!\!
\frac{[\a_i\pl\a_j\pl1]_1}{[\a_i\pl\a_j]_1}\,,\;\;n\mbox{ even.}\ea\right.\ee
For each set $(a,b,c,d,a)\in\G^4$ of spin states adjacent around a square face,
there exists a unique assignment of $i$ or of $i$ and $j$ in
exactly one of the five equations  of~\er{W} which defines the Boltzmann
weight $\Wsm{b\:\;}{c}{d}{a\:\;}{u}$.
The two forms of $\Wsm{a\pl\e_i}{a}{a\pl\e_i}{a}{u}$ in~\er{W} are equivalent,
except if $[2\a_i\pl1]_1=0$ (which occurs for example for $n$ even with $i=0$) in which case
only the second form can be used,
or if $[2\a_i\pl1\mi2\l]_1=0$ in which case only the first form can be used.
The equivalence of these two forms can be proved using the same method as that used
in \cite{JimMiwOka88b} to prove a similar equivalence for
the Boltzmann weights of that paper.

The Boltzmann weights of~\er{W} satisfy the Yang-Baxter equation,
\setlength{\unitlength}{7mm}
\be\el{YBE}\ba{c}\ru{3.5}\ds
\sum_{\stackrel{\ru{0.6}\ss g\:\in\:\B}{\sss
(G_{ag}G_{gc}G_{ge}=1)}}\!\!\!\!
\W{b}{c}{g}{a}{u}\,\W{c}{d}{e}{g}{u\pl v}\,\W{g}{e}{f}{a}{v}\;=\\
\ds\qquad\qquad\qquad\qquad
\sum_{\stackrel{\ru{0.6}\ss g\:\in\:\B}{\sss
(G_{bg}G_{fg}G_{gd}=1)}}\!\!\!\!
\W{c}{d}{g}{b}{v}\,\W{b}{g}{f}{a}{u\pl v}\,\W{g}{d}{e}{f}{u}\\
\text{0.55}{3}{0}{1.4}{l}{\stackrel{\ss\sum}{\sss g}}
\bpic(4,3.6)\multiput(0.5,1.5)(1,-1){2}{\line(1,1){1}}
\multiput(0.5,1.5)(1,1){2}{\line(1,-1){1}}
\multiput(2.5,0.5)(0,1){3}{\line(1,0){1}}
\multiput(2.5,0.5)(1,0){2}{\line(0,1){2}}
\put(1.5,1.5){\pp{}{u}}\put(3.05,1){\pp{}{u+v}}\put(3,2){\pp{}{v}}
\put(0.4,1.5){\pp{r}{b}}\put(1.5,0.4){\pp{t}{c}}
\put(2.5,0.4){\pp{t}{c}}\put(3.55,0.45){\pp{tl}{d}}
\put(3.6,1.5){\pp{l}{e}}\put(3.55,2.55){\pp{bl}{f}}
\put(1.5,2.6){\pp{b}{a}}\put(2.5,2.6){\pp{b}{a}}
\put(2.32,1.45){\pp{r}{g}}
\multiput(1.75,0.5)(0.25,0){3}{\pp{}{.}}
\multiput(1.75,2.5)(0.25,0){3}{\pp{}{.}}\epic
\text{1.3}{3}{0.6}{1.5}{}{=}
\text{0.6}{3}{0.3}{1.4}{}{\stackrel{\ss\sum}{\sss g}}
\bpic(4,3)\multiput(1.5,1.5)(1,-1){2}{\line(1,1){1}}
\multiput(1.5,1.5)(1,1){2}{\line(1,-1){1}}
\multiput(0.5,0.5)(0,1){3}{\line(1,0){1}}
\multiput(0.5,0.5)(1,0){2}{\line(0,1){2}}
\put(2.5,1.5){\pp{}{u}}\put(1,1){\pp{}{v}}\put(1.05,2){\pp{}{u+v}}
\put(0.4,1.5){\pp{r}{b}}\put(0.45,0.45){\pp{tr}{c}}
\put(1.5,0.4){\pp{t}{d}}\put(2.5,0.4){\pp{t}{d}}
\put(3.6,1.5){\pp{l}{e}}\put(1.5,2.6){\pp{b}{f}}
\put(2.5,2.6){\pp{b}{f}}\put(0.45,2.55){\pp{br}{a}}
\put(1.72,1.5){\pp{l}{g}}
\multiput(1.75,0.5)(0.25,0){3}{\pp{}{.}}
\multiput(1.75,2.5)(0.25,0){3}{\pp{}{.}}
\put(4.2,1.3){\p{}{,}}\epic\ea\ee
for all $(a,b,c,d,e,f,a)\in\G^6$ and $u,v\in\C$,
and the inversion relation
\setlength{\unitlength}{5mm}
\be\el{IR}\ba{c}\ds
\sum_{\stackrel{\ru{0.5}\ss e\:\in\:\B}{\sss(G_{ae}G_{ec}=1)}}\!\!
\W{b}{c}{e}{a}{\!\mi u}\;\W{e}{c}{d}{a}{u}\:=\:\rho(u)\:\delta_{bd}\\
\text{1}{3}{0}{1.4}{l}{\stackrel{\ss\sum}{\sss e}}
\bpic(5,3)\multiput(0.5,1.5)(3,-1){2}{\line(1,1){1}}
\multiput(0.5,1.5)(3,1){2}{\line(1,-1){1}}
\put(1.5,2.5){\line(1,-1){2}}\put(1.5,0.5){\line(1,1){2}}
\multiput(1.72,0.5)(0.26,0){7}{\pp{}{.}}
\multiput(1.72,2.5)(0.26,0){7}{\pp{}{.}}
\put(1.35,1.5){\pp{}{-u}}\put(3.5,1.5){\pp{}{u}}
\put(0.4,1.5){\pp{r}{b}}\multiput(1.5,0.4)(2,0){2}{\pp{t}{c}}
\put(4.6,1.5){\pp{l}{d}}\multiput(1.5,2.6)(2,0){2}{\pp{b}{a}}
\put(2.51,1.3){\pp{t}{e}}\epic
\text{5}{3}{5}{1.5}{r}{=\:\rho(u)\:\delta_{bd}\,,}\ea\ee
for all $(a,b,c,d,a)\in\G^4$ and $u\in\C$, where
\be\rho(u)\,=\,\frac{[\l\mi u]_4\,[\l\pl u]_4\,[1\mi u]_1\,[1\pl u]_1}
{[\l]_4^{\:\;2}\,[1]_1^{\:\;2}}\,.\ee
That the Yang-Baxter equation and inversion relation are
satisfied by the conjugate $A_{n,g}$ weights
of \er{W} can be proved using an approach similar to that used
in~\cite{JimMiwOka88b} to prove that these equations are satisfied by
the $B$ and $C$ weights of that paper. Although this constitutes a long
proof, it is the only one currently known to us.

It is straightforward to check that the weights of \er{W} also satisfy
an initial condition,
\be\el{IC}\W{b}{c}{d}{a}{0}\;=\;\delta_{bd}\,,\ee
reflection symmetry,
\be\el{RS}\W{b}{c}{d}{a}{u}\;=\;\W{d}{c}{b}{a}{u}\;=\;\W{b}{a}{d}{c}{u},\ee
crossing symmetry\footnote{\rule{0ex}{5ex}Explicitly, $\Wsm{b\;\,}{c}{d}{a\;\,}{u}=
\ds e^{\frac{i\pi}{g}(\l+\frac{\tau g}{2}-2u)}\,
\frac{\kappa_b\,\kappa_d}{\kappa_a\,\kappa_c}\;
\Wsm{c\;\,}{b}{a}{d\;\,}{\l\pl\frac{\tau g}{2}\mi u}$, where $q=e^{i\pi\tau}$ and\ru{4}\\
$\kappa_a=\left\{\ba{l}\ru{3}
\psi_{\!a}^{\,\hf}\;e^{\sss\mi\frac{i\pi}{2g}\!\!\sum_{j=\!1}^{\!\frac{n\pl1}{2}}\!\ss\a_j^2}
\;\,i^{\sss\sum_{j=\!1}^{\!\frac{n\pl1}{2}}\ss\a_j}
,\;\;n\mbox{ odd}\\
\ru{2}\psi_{\!a}^{\,\hf}\;e^{\sss\mi\frac{i\pi}{2g}\!\!\sum_{j=\!1}^{\!\frac{n}{2}}\!\ss\a_j^2}
,\;\;n\mbox{ even.}\ea\right.$},
and
graph symmetry for $n$ odd,
\be\W{b}{c}{d}{a}{u}\;=\;\W{Jb}{Jc}{Jd}{Ja}{u}\quad(n\mbox{ odd})\,,\ee
with $J$ as given in \er{J}. The weights are not invariant
under $J$ for $n$ even, $g$ even.

In Section 4, it will be useful to consider the Boltzmann weights as
face transfer matrices.  These are matrices $X_1(u),\ldots,X_r(u)$, for some fixed
$r\in\NN$, whose rows and columns are labeled by elements of $\G^{r\pl1}$
and whose entries are defined by
\be\el{X}X_j(u)_{(a_0,\ldots,a_{r\pl1}),(b_0,\ldots,b_{r\pl1})}\;=\;
\W{a_j}{a_{j\pl1}}{b_j}{a_{j\mi1}}{u}\,\prod_{k=0\atop k\ne j}^{r\pl1}\delta_{a_k b_k}\,.\ee
It follows from the Yang-Baxter equation~\er{YBE} that these matrices satisfy
\be\el{XYBE}X_j(u)\;X_{j+1}(u\pl v)\;X_j(v)\;=\;X_{j+1}(v)\;X_j(u\pl v)\;X_{j+1}(u)\,,\ee
and from the inversion relation~\er{IR} that they satisfy
\be\el{XIR}X_j(-u)\:X_j(u)\;=\;\rho(u)\:I\,.\ee
It is also seen immediately from~\er{X} that
\be\el{XC}X_j(u)\:X_k(v)\;=\;X_k(v)\:X_j(u)\,,\quad|j\mi k|\ge2\,.\ee

\sect{The Critical Case}
In this section we consider the Boltzmann weights of Section~3 at $q=0$,
where $q$ is the elliptic nome of~\er{ef}, this being expected to correspond
to criticality.
In particular, we examine the
various general properties outlined in~\cite{DifZub90a,DifZub90b,Dif92}.

For $q=0$ we now have
\be[u]_1\,=\,\sin\Bigl(\frac{u\pi}{g}\Bigr)\equiv s(u)\,,\qquad[u]_4\,=\,1\,,\ee
throughout \er{W}--\er{psiai2}.
It can be shown, incidentally, that
$\psi_a$ are now entries of the Perron-Frobenius
eigenvector of $G$,
\be\sum_{b\:\in\:\B}G_{ab}\:\psi_b\;=\;\frac{s(n\pl1)}{s(1)}\;\psi_a\,.\ee
Defining
\be U_j\;=\;X_j(1)\,,\ee
it follows using standard trigonometric identities
on $\Wsm{a\pl\e_i\;\;}{a\pl\e_i\pl\e_j}{a\pl\e_i}{a}{u}$ and
$\Wsm{a\pl\e_i}{a}{a\pl\e_i}{a}{u}$
in~\er{W}, that the critical face transfer matrices can be decomposed as
\be\el{XIU}X_j(u)\;=\;\frac{s(1\mi u)}{s(1)}\:I\;+\;\frac{s(u)}{s(1)}\:U_j\,.\ee
Substituting this into \er{XYBE}--\er{XC}, it then follows
that $U_j$ form a matrix representation of
the Hecke algebra,
\be\el{HA}\ba{r@{\;\,}c@{\;\,}l}\ru{2.4}U_j^2&=&2\cos(\pi\!/\!g)\:U_j\\
\ru{2.4}
U_j\:U_{j+1}\:U_j-U_j&=&U_{j+1}\:U_j\:U_{j+1}-U_{j+1}\\
\ru{0}U_j\:U_k&=&U_k\:U_j\,,\quad|j\mi k|\ge2\,.\ea\ee
The critical Boltzmann weights are also expected to
satisfy the trace property,
\be\el{TP}\frac{1}{2\cos(\pi\!/\!g)}
\sum_{\stackrel{\ru{0.6}\ss c\:\in\:\B}{\sss(G_{ac}G_{cb}=1)}}\!
\W{c}{b}{c}{a}{1}\:=\:\N_{(1,2,1,\ldots,1)\,ab}\,,\ee
for all $a,b\in\B$ with $(G^2)_{ab}\ne0$, the
Markov property,
\be\el{MP}\sum_{\stackrel{\ru{0.6}\ss c\:\in\:\B}{\sss(G_{ac}=1)}}
\W{a}{c}{a}{b}{u}\psi_c\:=\:\frac{s(n\pl1\mi u)}{s(1)}\,\psi_a\,,\ee
for all $(a,b)\in\G^1$ and $u\in\C$,
and the quantum group antisymmetrizer property,
\be\el{ASP}\sum_{\s\in S_{n\pl2}}\;\prod_{\tau_j\in T_{\s}}\!Y_j\,=\,0,\ee
where $Y_j\,=\,e^{\frac{i\pi}{g}}U_j-I$,
$S_n$ is the symmetric group on $n$ objects, $1,2,\ldots,n$, and
$T_\s$ is any minimal-length set of
transpositions $\tau_j\equiv(j\leftrightarrow j\pl1)$
whose product in a particular order is the permutation $\s$, with the product in~\er{ASP}
taken in the same order.  (If $\s$ is the identity permutation,
then $T_{\s}=\emptyset$ and $\prod_{\tau_j\in T_{\s}}Y_j$ is taken as the identity matrix.
Also, it follows from~\er{HA} that $Y_j$ satisfy
the braid relations $Y_j\,Y_{j\pl1}\,Y_j=Y_{j\pl1}\,Y_j\,Y_{j\pl1}$ and $Y_j\,Y_k=Y_k\,Y_j$
for $|j\mi k|\ge2$, which implies that
$\prod_{\tau_j\in T_{\s}}\!Y_j$ is independent of the choice of $T_\s$ for given $\s$.)
The imposition of the quantum group antisymmetrizer property~\er{ASP} on the Hecke algebra~\er{HA}
can be regarded as taking a quotient of that algebra.

We have checked that properties \er{TP}--\er{ASP} are
satisfied by the critical conjugate $A_{n,g}$ Boltzmann weights
for a variety of choices of $n$ and $g$,
and we believe that these properties hold for all $n$ and $g$.
The significance of these properties is
discussed in \cite{DifZub90a,DifZub90b,Dif92}.

\sect{Discussion}
We have presented Boltzmann weights for integrable lattice models based on fundamental
conjugate $A_{n,g}$ nimrep graphs.
We derived these weights through direct and detailed consideration of the Yang-Baxter equation
as applied to these graphs, but it would clearly be of interest to obtain an alternative
systematic algebraic procedure for generating the weights
which reflects one or more of the procedures used
in~\cite{PetZub02,GabGan02} to generate the nimreps.
Such a procedure would be expected to reveal relationships between the conjugate $A_{n,g}$ Boltzmann weights
and the $B^{(1)}_l$, $C^{(1)}_l$ and $A^{(2)}_l$
Boltzmann weights of~\cite{JimMiwOka88b,Kun91},
reflecting such relationships at the level of the nimreps.

It is also expected, at least at criticality, that there exist cells which intertwine
the conjugate $A_{n,g}$ Boltzmann weights with the
identity $A_{n,g}$ Boltzmann weights of~\cite{JimMiwOka87a,JimMiwOka88a}.
If such cells were obtained and shown to satisfy intertwining and inversion equations
(specifically (4.6a,b,c) of~\cite{DifZub90a}),
it would then follow immediately that key relations known to be satisfied
by the identity $A_{n,g}$ Boltzmann weights, including the Yang-Baxter equation and the
quantum group antisymmetrizer property, are also satisfied by the conjugate $A_{n,g}$ weights.

The lattice models given here are related to the fundamental vector representation
$(2,1,\ldots,1)$ of $A_{n,g}$, with horizontal and vertical adjacency conditions on the lattice
both specified by $\N_{(2,1,\ldots,1)}$.  In the case of the identity $A_{n,g}$
lattice models of~\cite{JimMiwOka87a,JimMiwOka88a}, an $A_{n,g}$
fusion procedure was applied in~\cite{JimMiwOka87b,JimMiwOka88c,JimKunMiwOka88}
to the fundamental vector Boltzmann weights to construct weights
which still satisfy the Yang-Baxter equation and have
horizontal and vertical adjacency conditions given by $N_{\l}$ and
$N_{\mu}$ for any $\l,\mu\in P_{n,g}$.
(In fact the graphs can then contain multiple edges between two given vertices, which requires
the introduction of additional spins on the edges of the lattice.)
And it is expected, at least at criticality, that essentially the same
fusion procedure can be applied to the conjugate $A_{n,g}$ fundamental vector
weights to construct conjugate $A_{n,g}$
weights with horizontal and vertical adjacency conditions given by arbitrary $\N_{\l}$ and
$\N_{\mu}$.  An important component
of this procedure is the construction of projection matrices $Q_{\l}(a,b)$, for
each $\l\in P_{n,g}$ and $a,b\in\B$.  The rows and columns of $Q_{\l}(a,b)$ are
labeled by the $r$-step paths on $\G_{n,g}$ between $a$ and $b$, where $r$ is the number of
boxes in the Young diagram corresponding to $\l$, $r=\sum_{i=1}^n i(\l_i\mi1)$, and the rank
of the projector $Q_{\l}(a,b)$ is $\N_{\l\,ab}$.  (For the case $\l=(1,2,1,\ldots,1)$,
this projector is given by $Q_{\l}(a,b)_{(a,c,b),(a,d,b)}=\Wsm{c\;\,}{b}{d}{a\;\,}{1}\Big/(2\cos(\pi/g))$
and the trace property~\er{TP} is the equation for its rank, since the rank of a projector is
simply its trace). If it can be shown that the $A_{n,g}$ fusion procedure
of~\cite{JimMiwOka87b,JimMiwOka88c,JimKunMiwOka88} is indeed applicable
to the critical conjugate $A_{n,g}$ Boltzmann weights, then an immediate corollary would be
the nonnegativity of each $\N_{\l\,ab}$ for conjugate $A_{n,g}$
(since these integers can then be interpreted as ranks of projectors), this nonnegativity
not having been proved in general in~\cite{PetZub02,GabGan02}.

We also note that an approach to the generation of modular invariants and nimreps,
in which the nimrep matrices are guaranteed to have nonnegative entries,
is that of $\alpha$-induction, as described in~\cite{BocEvaKaw99,BocEvaKaw00}.  If this
were applied to the case of the $A_{n,g}$ conjugate modular invariant, the
critical conjugate $A_{n,g}$ Boltzmann weights would be used together with intertwiner cells
to construct braided subfactors which, through $\alpha$-induction, would
be expected to lead to the nimreps of~\cite{PetZub02,GabGan02}.

For a few special cases of $n$ and $g$, other integrable
interaction-round-a-face lattice models are known which are also based on the
fundamental vector conjugate $A_{n,g}$ graphs.
For example, it can be seen in the figures in Section 2 that
$\G_{n,n\pl3}$, $n$ odd, corresponds to a $D$ Dynkin diagram,
with other models based on these having been given in~\cite{Pas87b}, and
that $\G_{2,g}$, $g$ even, and $\G_{n,n\pl4}$, $n$ even, both correspond to `dilute' $A$
Dynkin diagrams, for which other models were given in~\cite{WarNieSea92},
but these other models differ intrinsically from the conjugate $A_{n,g}$ models.
In particular, the critical Boltzmann weights of~\cite{Pas87b}
lead to relations of the type \er{TP}--\er{ASP}, but the data in these relations then pertains
to $A_{1,n\pl3}$ $D_{\!\frac{n+5}{2}}$ nimreps as distinct from $A_{n,n\pl3}$
conjugate nimreps, while the critical Boltzmann weights of~\cite{WarNieSea92}
lead to representations of the dilute Temperley-Lieb algebra as opposed to the
Hecke algebra.  (A further, but trivial, example is
$\G_{1,g}$, which corresponds to the $A_{g-1}$ Dynkin diagram.
Models based on these graphs were given in~\cite{AndBaxFor84} and as
the $n=1$ case of the identity $A_{n,g}$ models of~\cite{JimMiwOka87a,JimMiwOka88a},
and these are equivalent to the conjugate $A_{1,g}$ models,
but this corresponds to the fact that $A_{1,g}$ conjugation is simply the identity,
so that all objects related to the conjugate
are equivalent to those already known for the identity.)
When deriving the Boltzmann weights for conjugate $A_{n,g}$, we
observed in various cases
that by imposing the properties of Section~4 at criticality, the solution
to the Yang-Baxter equation
was unique, up to unimportant normalisation, gauge transformation,
shifts in the spectral parameter or relabeling of the graph.
We thus conjecture that for any $n$ and $g$, the Boltzmann
weights given here constitute an essentially unique
solution of the Yang-Baxter equation
based on $\G_{n,g}$ which at criticality satisfies the
properties of Section~4 for conjugate $A_{n,g}$.

\end{document}